\documentclass[a4paper, 11pt]{article}
\pdfoutput=1

\usepackage{jheppub}
\usepackage{afterpage}
\usepackage{ifthen}
\usepackage{mciteplus}

\newcommand{\Sla}[1]{/\!\!\!\!#1}
\def\ie{{\it i.e. }}
\def\lsim{\raise0.3ex\hbox{$\;<$\kern-0.75em\raise-1.1ex\hbox{$\sim\;$}}}
\def\gsim{\raise0.3ex\hbox{$\;>$\kern-0.75em\raise-1.1ex\hbox{$\sim\;$}}}

\title{LHC Run I Bounds on Minimal Lepton Flavour Violation in Type--III
  See--saw: A Case Study}

\author[a]{Nuno Rosa Agostinho}
\emailAdd{nunorosaagostinho@ub.edu}

\author[b]{O.\ J.\ P.\ \'Eboli}
\emailAdd{eboli@if.usp.br}

\author[a,c,d]{M.~C.~Gonzalez-Garcia,}
\emailAdd{maria.gonzalez-garcia@stonybrook.edu}

\affiliation[a] {Departament de Fis\'{\i}ca Qu\`antica i
  Astrof\'{\i}sica and Institut de Ciencies del Cosmos, Universitat de
  Barcelona, Diagonal 647, E-08028 Barcelona, Spain}

\affiliation[b]{Instituto de F\'{\i}sica, 
Universidade de S\~ao Paulo, S\~ao Paulo -- SP, Brazil.}

\affiliation[c]{Instituci\'o Catalana de Recerca i Estudis
  Avan\c{c}ats (ICREA), Pg. Lluis Companys 23, 08010 Barcelona,
  Spain.}

\affiliation[d]{C.N.~Yang Institute for Theoretical Physics, State
  University of New York at Stony Brook, Stony Brook, NY 11794-3840,
  USA}

\abstract{
  We study the bounds on minimal lepton flavour violation in the context of
  Type-III see--saw imposed by LHC Run I search for events which contain two
  charged leptons (either electron or muons of equal or opposite
  sign), two jets from a hadronically decaying $W$ boson and large
  missing transverse momentum.  In this scenario the flavour structure of
  the couplings of the triplet fermions to the Standard Model leptons
  can be reconstructed from the neutrino mass matrix and lepton number
  violation is very suppressed. We find that using the information on
  charge and flavour of the leptons in the above final state it is possible
  to unambiguously rule out this scenario with triplet masses lighter
  than 300 GeV at 95\% CL. The same analysis allows to exclude triplet
  masses masses up to 480 GeV at 95\% CL for normal ordering of
  neutrino masses and specific values of
  a Majorana CP phase currently undetermined by neutrino physics.}

\preprint{YITP-SB-17-31}

\begin{document}

\maketitle

\section{Introduction}

Massive neutrinos are our first and only undoubted evidence of physics
beyond the Standard Model (SM).  The evidence arises from a variety of
neutrino experiments which have detected the effect of their mass via
the flavour oscillation of neutrinos with energies ranging between
tens of keV and tens of GeV \cite{GonzalezGarcia:2007ib}. The obvious
question that this observation raises is that of the dynamics of the
New Physics (NP) responsible for the neutrino mass.

It is well known that in the framework of effective operators for NP
there is just one dimension-five operator which can be
built~\cite{Weinberg:1979sa},
${\cal O}=\alpha_5 /\Lambda_{LN}\,
L_L^{} L_L^{} HH$, where $L_L^{}$ and $H$ are the leptonic and Higgs
$SU(2)_L$ doublets.  This operator breaks total lepton number and
after electroweak symmetry breaking it generates Majorana masses for
the neutrinos $m_\nu \sim \alpha_5 v^2/\Lambda_{LN}$, where $v$ is the
SM Higgs vacuum expectation value (vev). This explains the lightness
of the neutrino mass due to the large scale of total lepton number
violation $\Lambda_{LN}$.  In the simplest UV completions, this
dimension-5 operator can be generated by the tree-level exchange of
three types of new states: lepton singlets in the Type--I see-saw
scenarios~\cite{Minkowski:1977sc, Yanagida:1979as, GellMann:1980vs,
  Mohapatra:1979ia}, a scalar triplet for the Type--II
mechanisms~\cite{Konetschny:1977bn, Cheng:1980qt, Lazarides:1980nt,
  Schechter:1980gr, Mohapatra:1980yp}, and lepton triplets for
Type--III models~\cite{Foot:1988aq}.  In any of these mechanisms the
smallness of the neutrino mass can be naturally explained with Yukawa
couplings $\lambda\sim{\cal O}(1)$ if the masses of the new states are
$M\sim \Lambda_{LN}\sim 10^{14-15}$ GeV.  Notwithstanding consistent
models of lower scale see--saw exist in the literature for some time;
see {\em e.g.} \cite{Mohapatra:1986bd,Kersten:2007vk,Chen:2011de}.

Given the energy range of the neutrino experiments it is clear that if
the NP scale is beyond $\sim$ GeV it is not possible to clarify its
origin within the oscillation neutrino experiments themselves.  On the
other hand, the high energy frontier is currently being explored by the
CERN Large Hadron Collider (LHC) which has been running for eight
years with a reach to NP at the TeV scale. So far no clear evidence of
NP has been observed at LHC which bears implications for models
constructed to explain the neutrino masses containing new states at
the TeV scale.

Generically the expected event rates at LHC in neutrino mass models
depend not only on the mass and weak charge of the new states
involved, but also on the flavour structure of their couplings which
determines their decay modes. A priory, the decay channels are
arbitrary  in most Type--I and Type--III see--saw
models, making difficult to derive unambiguous constraints on the NP
scale in these scenarios~\cite{delAguila:2008cj,
  Aguilar-Saavedra:2013twa}.  For example searches for triplet leptons
of the Type--III see--saw models have been performed both by CMS
\cite{CMS:2012ra,CMS:2017wua, CMS:2016hmk,CMS:2015mza} and ATLAS
\cite{Aad:2015cxa,ATLAS:2013hma} collaborations, however, most of
these searches have been carried out within the context of simplified
models such as Ref.~\cite{Biggio:2011ja} and the derived bounds can be
evaded depending on which is the dominant decay mode of the triplet
leptons.

One exception are see--saw models which extend the principle of
minimal flavour violation to the leptonic sector.  Minimal flavour
violation was first introduced for quarks~\cite{Chivukula:1987py,
Buras:2000dm, DAmbrosio:2002vsn} as a way to explain the absence of
NP effects in flavour changing processes in meson decays. The basic
assumption is that the only source of flavour mixing in the full
theory is the same as in the SM, \ie the quark Yukawa couplings.
This idea was latter on extended for leptons~\cite{Cirigliano:2005ck,
Davidson:2006bd} and in particular to TeV scale see--saw
models~\cite{Cirigliano:2005ck, Davidson:2006bd, Gavela:2009cd,Alonso:2011jd}. 

From the point of view of LHC phenomenology minimal lepton flavour
violation (MLFV) see--saw models are attractive since, a) the new
states can be light enough to be produced at LHC, and b) their
signatures are fully determined by the neutrino parameters.  As
discussed in Ref.~\cite{Gavela:2009cd} scalar (Type-II) see--saw with
light doublet-triplet mixing is a light scale MLFV model by
construction (for early study of their observability see for example
~\cite{Perez:2008ha,Garayoa:2007fw}).
In Ref.~\cite{Gavela:2009cd}  simple MLFV models for fermionic
see--saw were also presented. In Type--I see--saw the new states are
singlets under the SM gauge group, and therefore, they can only be
produced via their mixing with the SM neutrinos. This leads to small
production rates which makes the model only marginally testable at
LHC.  Type-III see--saw fermions, on the contrary, are $SU(2)_L$
triplets with weak-interaction pair-production cross section, and
consequently, having the potential to allow for tests of the
hypothesis of MLFV.

In Ref.~\cite{Eboli:2011ia} we studied the potential of LHC to unravel
the existence of triplet fermionic states that appear in MLFV Type-III
see--saw models of neutrino mass. Here, we obtain the bounds on the NP
scale of this scenario that originates from the
ATLAS~\cite{Aad:2015cxa} searches in the final state topology
containing two charged leptons (electrons and/or muons), two jets
compatible with a hadronically decaying $W$ and missing transverse
momentum.  This ATLAS analysis is best suited for testing the MLFV
scenario because it classifies the final states in the different
flavour and charge combinations, allowing to fully exploit the
predictions of the MLFV Type-III see--saw model.  We find that using
this ATLAS data it is possible to unambiguously rule out the MLFV
Type--III see--saw scenario with triplet masses lighter than 300 GeV
at 95\% CL irrespective of the neutrino mass ordering and other
unknowns in the light neutrino sector, in particular of the so--far
undetermined Majorana CP violating phase.  The same analysis allows to
rule out triplet masses masses up to 480 GeV at 95\% CL for normal
order (NO) scenarios and specific values of that phase.

This paper is organized as follows.  We first summarize in
Sec.~\ref{sec:model} the basics of the MLFV Type-II see--saw model and
we quantify the allowed range of the relevant couplings controlling
the triplet decay modes as derived from the present analysis of
neutrino oscillation data from Ref.~\cite{Esteban:2016qun}. Section
~\ref{sec:simulation} describes our simulation of signal events by the
reaction $pp\rightarrow l l' j j \nu\nu$ with $ l^{(\prime)} = e$ or
$\mu$ in the context of the MLFV Type-III see--saw model.  The
quantification of the bounds is presented in Sec.~\ref{sec:results}.
In doing so we have put special emphasis and we have quantified how
the information on flavour and charge information of the produced
leptons is important for maximal sensitivity to MLFV.

\section{MLFV Type--III see--saw model}
\label{sec:model}

In Ref.~\cite{Eboli:2011ia} we introduced the simplest MLFV Type-III
see--saw model which was adapted from the Type-I one presented in
Ref.~\cite{Gavela:2009cd}. For completeness we summarize here its main
features.

The particle contents of model is that of the SM extended with two
fermion triplets
$\vec{\Sigma} = \left( \Sigma_1,\Sigma_2,\Sigma_3\right) $ and
$\vec{\Sigma}^\prime = \left( \Sigma'_1,\Sigma'_2,\Sigma'_3\right)$,
each one formed by three right-handed Weyl spinors of zero
hypercharge. Hence, the Lagrangian is
\begin{eqnarray}
 {\mathcal L}={\mathcal L}_{SM}+{\mathcal L}_K+{\mathcal
 L}_Y+{\mathcal L}_\Lambda
\label{eq:lag}
\end{eqnarray}
with
\begin{eqnarray}
{\mathcal
L}_K=&&i\left(\overline{\vec{\Sigma}}\Sla{D}_\mu\vec{\Sigma}+
\overline{\vec{\Sigma}^\prime}\Sla{D}_
\mu\vec{\Sigma}^\prime\right)
\;,\\ 
 {\mathcal
L}_Y=&&-Y_i^\dag\overline{L^w_{L
i}}\left(\vec{\Sigma}\cdot\vec{\tau}\right)\tilde{\phi}- \epsilon
Y_i^{\prime\dag}\overline{L^w_{L
i}}\left(\vec{\Sigma}^\prime\cdot\vec{\tau}\right)\tilde{\phi}+h.c.
\;, \\
{\mathcal
L}_\Lambda=&&
-\frac{\Lambda}{2}\left(\overline{\vec{\Sigma}^c}\vec{\Sigma}^\prime
+\overline{\vec{\Sigma}^{\prime c}}\vec{\Sigma}\right)
-\frac{\mu}{2}\overline{\vec{\Sigma}^{\prime c}}\vec{\Sigma}^\prime
-\frac{\mu'}{2}\overline{\vec{\Sigma}^c}\vec{\Sigma}
+h.c. \;.
\end{eqnarray}
Here $\vec{\tau}$ are the Pauli matrices, the gauge covariant
derivative is defined as
$D_\mu=\partial_\mu+ig\vec{T}\cdot\vec{W}_\mu$, where $\vec{T}$ are
the three-dimensional representation of the $SU(2)_L$ generators,
$\phi$ stands for the SM Higgs doublet field, and
$L^w_{i}=(\nu^w_i,\ell^w_i)^T$ are the three weak eigenstate lepton
doublets of the SM. The parameters $\epsilon$, $\mu$ and $\mu'$ are
flavour-blind and {\sl small}, {\em i.e.}, the scales $\mu$ and $\mu'$ are
much smaller than $\Lambda$ and $v$ while $\epsilon\ll 1$.

The Lagrangian in Eq.~\eqref{eq:lag} breaks total lepton number due to
the simultaneous presence of the Yukawa terms $Y_i$ and
$\epsilon Y'_i$ as well as to the existence of the $\mu$ and $\mu'$
terms. Thus in the limit $\mu,\mu',\epsilon\rightarrow 0$ it is
possible to define a conserved total lepton number by assigning
$L(L^w)=L(\Sigma)=-L(\Sigma^\prime)=1$. Without any loss of generality
one can work in a basis in which $\Lambda$ is real while both $Y$ and
$Y'$ are complex. In general the parameters $\mu$ and $\mu'$ would be
complex, but for the sake of simplicity we take them to be real in
what follows though it is straight forward to generalize the
expression to include the relevant phases~\cite{Abada:2007ux}.

  \begin{figure}[h!]
    \centering

\includegraphics[width=0.7\textwidth]{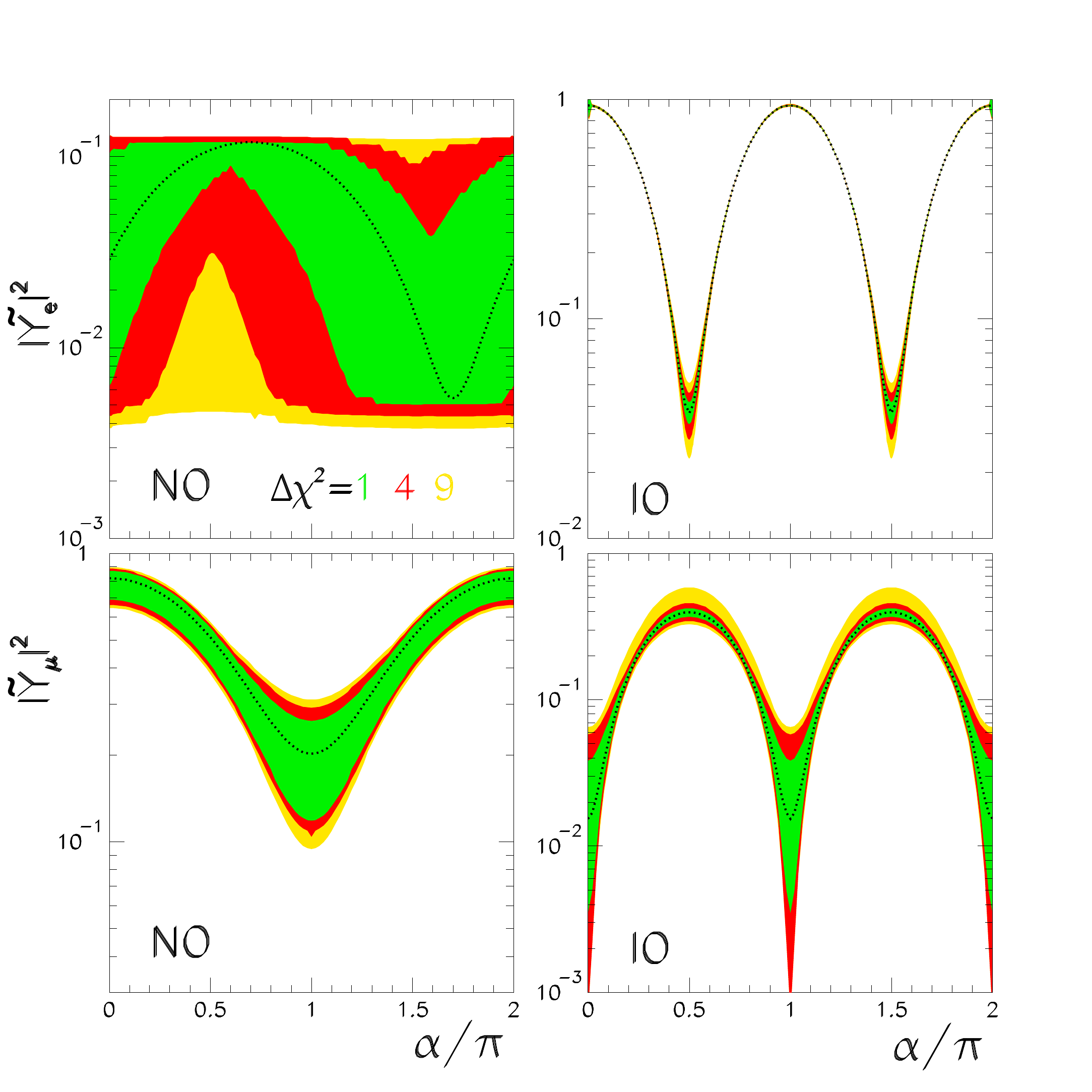}
\includegraphics[width=0.7\textwidth]{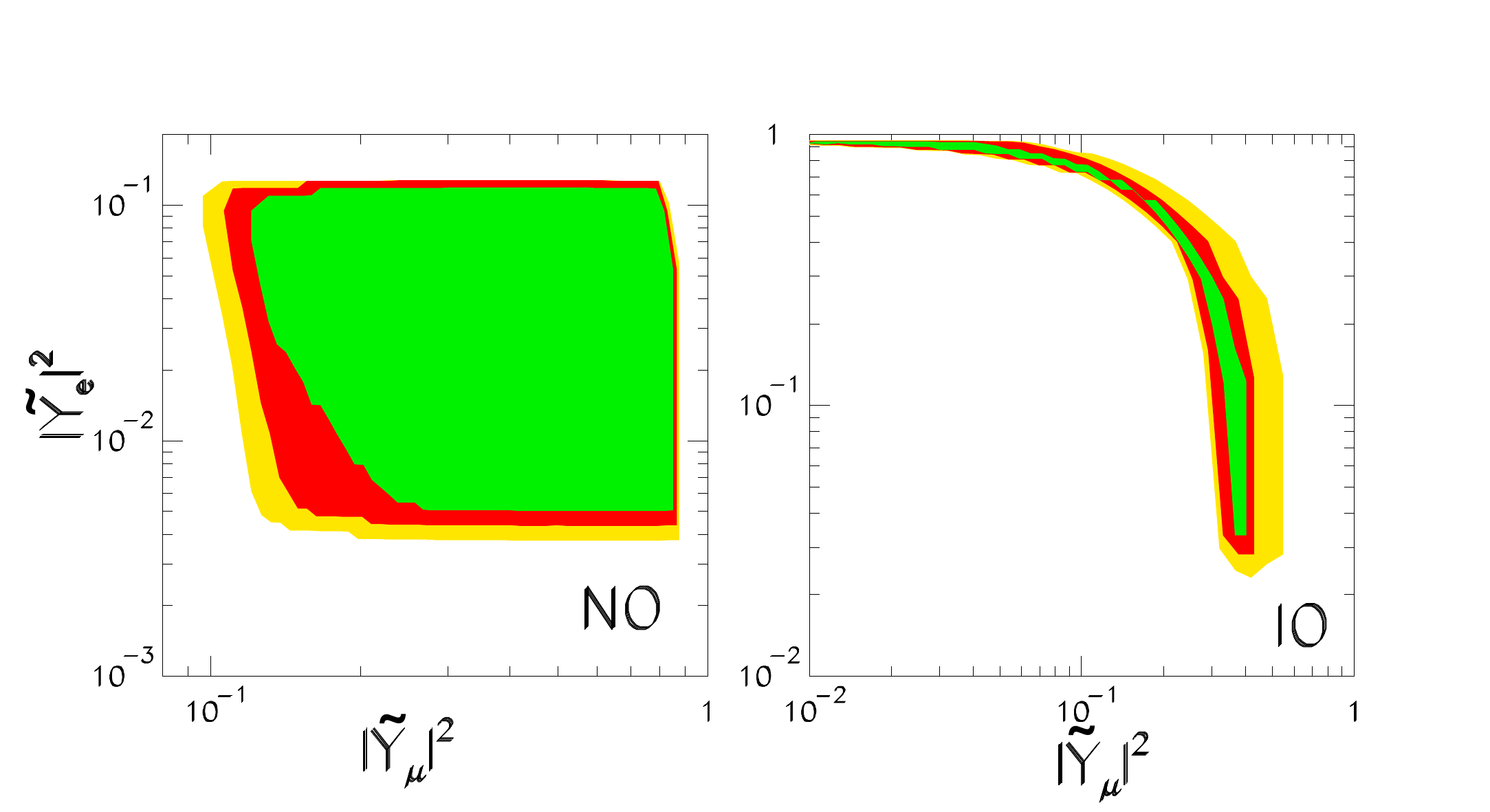}
\caption{ Allowed ranges of the Yukawa couplings $|\tilde Y_e|^2\equiv
  |Y_1|^2/y^2$ and $|\tilde Y_\mu|^2\equiv |Y_2|^2/y^2$ obtained from
  the global analysis of neutrino data in Ref.~\cite{Esteban:2016qun}.
  The upper four
  panel shows the values of the couplings as a function of the unknown
  Majorana phase $\alpha$. The correlation between the two couplings
  is shown in the two lower panels.  The left (right) panels correspond
  to normal (inverted) ordering. The dotted line corresponds to the
  best fit values. The ranges in the filled green, red and yellow areas are shown at 
  1$\sigma$, $2\sigma$, and 3$\sigma$ with 1 dof ($\Delta\chi^2=1,4,9$
  respectively). }
  \label{fig:yuk}
\end{figure}

After electroweak symmetry breaking, in the unitary gauge, the
leptonic mass matrices are
\begin{eqnarray}
{\mathcal L}_m
&=&-\frac{1}{2}\left(\overline{\vec{\nu^w_L}^c}\ 
\overline{\tilde N_R}\ \overline{\tilde N_R^{\prime}}\right)
M_0\left(\begin{array}{c}
       \vec{\nu^w_L}\\\tilde N_R^c\\\tilde N_R^{\prime c}
      \end{array}\right)
-\left(\overline{\vec{\ell^w_L}}\ \overline{E_L}\ 
\overline{E_L^{\prime}}\right)
M_\pm
\left(\begin{array}{c}\vec{\ell^w_R}\\E_R\\E_R^{\prime}\end{array}\right)
+h.c. 
\label{eq:lmass}
\end{eqnarray}
with
\begin{eqnarray}
M_0=\left(\begin{array}{ccc}
       0&\frac{v}{\sqrt{2}}Y^T&\epsilon\frac{v}{\sqrt{2}}Y^{\prime T} \\
  \frac{v}{\sqrt{2}}Y&\mu'&\Lambda \\
\epsilon\frac{v}{\sqrt{2}}Y^{\prime}&\Lambda&\mu
      \end{array}\right) 
&&
\;\;\;\;
\hbox{ and }
\;\;\;\;
M_\pm=
\left(\begin{array}{ccc}
\frac{v}{\sqrt{2}}Y^\ell&vY^\dagger&\epsilon vY^{\prime \dagger} \\
0&\mu'&\Lambda \\
0&\Lambda&\mu
\end{array}\right) \; ,
\end{eqnarray}
where $Y^\ell$ are the charged lepton Yukawa couplings of the SM.
$\vec\nu^w$ and $\vec\ell^w$ are column vectors containing
respectively the three neutrinos and charged leptons of the SM in the
weak basis. The charge eigenstates Dirac fermions $E$ and $E'$ and the
neutral Majorana fermions $\tilde N$ and $\tilde N^\prime$ are defined
in terms of the triplet states,
$\Sigma^{(\prime)}_\pm=\frac{1}{\sqrt{2}}\left(\Sigma^{(\prime)}_1\mp
  i\Sigma^{(\prime)}_2\right)$, and
$\Sigma^{(\prime)}_0=\Sigma^{(\prime)}_3$, as
\begin{eqnarray}
E^{(\prime)}
=\Sigma^{(\prime)}_-+{\Sigma_+^{(\prime)}}^c &\;\;\;\; 
\tilde N^{(\prime)}=\Sigma^{(\prime)}_0+{\Sigma^{(\prime)}_0}^c 
\; .
\end{eqnarray}

The mass basis is composed of:
\begin{itemize}
\item Three light Majorana neutrinos $\nu_i$ (with the
lightest one being massless) and three light charged leptons
leptons $\ell_i$ with masses 
\begin{eqnarray}
m^{diag}_\nu&=& 
{V^\nu}^T \left[-\frac{v^2}{2\Lambda} 
  \epsilon\left[\left(Y'-\frac{1}{\epsilon}\frac{\mu}{2\Lambda} Y\right)^T Y
    + Y^T \left(Y'-\frac{1}{\epsilon}\frac{\mu}{2\Lambda} Y\right)\right]\right]V^\nu  \; , \\
m^{diag}_\ell&=&
\frac{v}{\sqrt{2}}  {V^\ell}^\dagger_R Y^{\ell\dagger}
\left[1-\frac{v^2}{2\Lambda^2} 
Y^\dagger Y\right] V^\ell_L \; ,
\end{eqnarray}
where $V^\nu$ and $V^\ell_{L,R}$ being $3\times 3$ unitary matrices
and in general, one can choose the flavour basis such that
$ V^\ell_L= V^\ell_R=I$.
\item Two charged heavy leptons, $E_1^-$ and $E_2^+$,
both with masses $M\simeq\Lambda (1 \mp \frac{\mu+\mu'}{2\Lambda})$.
\item Two heavy Majorana neutral leptons and two charged
heavy leptons also with masses
$M\simeq\Lambda (1 \mp \frac{\mu+\mu'}{2\Lambda})$ with which we build
a quasi-Dirac heavy state $N$.
\end{itemize}
%


The relation between the weak and mass eigenstate to first order in
the small parameters $\mu$, $\mu'$ and $\epsilon$ is:
\begin{eqnarray}
\nu_L^w&=&V^\nu \nu_L 
+ \frac{v}{\sqrt{2}\Lambda}Y^\dagger N_L
+ \frac{v}{\sqrt{2}\Lambda}\left(\epsilon Y^{\prime \dagger}-
\left(\frac{3\mu+\mu^\prime}{4\Lambda}\right)Y^\dagger\right)N_R^c
\;, 
\\
\ell^w_L&=&\ell_L 
+\frac{v}{\Lambda}Y^\dagger E_{1L}^-
+\frac{v}{\Lambda}\left(\epsilon Y^{\prime \dagger}-\left(\frac{3\mu+\mu^\prime}{4\Lambda}\right)
Y^\dagger\right)E_{2R}^{+c}  \; , \\
\ell^w_R&=&\ell_R \; , \\
\tilde N_L&=&N_R^c-\left(\frac{\mu-\mu^\prime}{4\Lambda}\right)N_L
-\frac{v}{\sqrt{2}\Lambda}\left(\epsilon Y^{\prime}-\frac{\mu}{\Lambda}Y\right)V^\nu \nu_L\; , \\
\tilde N'_L&=&N_L+\left(\frac{\mu-\mu^\prime}{4\Lambda}\right)N_R^c
-\frac{v}{\sqrt{2}\Lambda}Y V^\nu \nu_L    \;,
\\
E_L&=&E_{2R}^{+c}
-\left(\frac{\mu-\mu^\prime}{4\Lambda}\right)E_{1L}^-
-\frac{v}{\Lambda}\left(\epsilon Y^{\prime}-\frac{\mu}{\Lambda}Y\right)\ell_L \; , \\
E_R&=&E_{1R}^-
-\left(\frac{\mu-\mu^\prime}{4\Lambda}\right)E_{2L}^{+c}\; , \\
E'_L&=&E_{1L}^-
+\left(\frac{\mu-\mu^\prime}{4\Lambda}\right)E_{2R}^{+c}
-\frac{v}{\Lambda}Y \ell_L  \; , \\
E'_R&=&E_{2L}^{+c}
+\left(\frac{\mu-\mu^\prime}{4\Lambda}\right)E_{1R}^- \; .
\end{eqnarray}
From the above relations it follows that the neutral weak interactions
of the light states take the same form as that on the SM and the
charged current interactions involve a 3x3 unitary matrix
$U_{LEP}=V^\nu$ which after phase redefinition of the light charged
leptons, can be chosen
\begin{eqnarray}
\!\!\!\!\!\!\!\!\!\!\!\!\!\!\!\!
U_{LEP}=&=&
    \begin{pmatrix}
	1 & 0 & 0 \\
	0 & c_{23}  & {s_{23}} \\
	0 & -s_{23} & {c_{23}}
    \end{pmatrix}
\!\!\!
    \begin{pmatrix}
	c_{13} & 0 & s_{13} e^{-i\delta_\text{CP}} \\
	0 & 1 & 0 \\
	-s_{13} e^{i\delta_\text{CP}} & 0 & c_{13}
    \end{pmatrix}
\!\!\! 
    \begin{pmatrix}
	c_{21} & s_{12} & 0 \\
	-s_{12} & c_{12} & 0 \\
	0 & 0 & 1
    \end{pmatrix}
\!\!\!
        \begin{pmatrix}
	  e^{-i \alpha}
          & 0 & 0 \\
	0 & e^{i \alpha} & 0 \\
	0 & 0 & 1
    \end{pmatrix},
\end{eqnarray}
where $c_{ij} \equiv \cos\theta_{ij}$ and
$s_{ij} \equiv \sin\theta_{ij}$.  The angles $\theta_{ij}$ can be
taken without loss of generality to lie in the first quadrant,
$\theta_{ij} \in [0,\pi/2]$ and the phases
$\delta_\text{CP},\; \alpha\in [0,2\pi]$.  The leptonic mixing matrix
contains only one Majorana phase because there are only two heavy
triplets and consequently only two light neutrinos are massive while
the lightest one remains massless. Also notice that unitarity
violation in the charged current and flavour mixing in the neutral
current of the light leptons in generated at higher
order~\cite{Schechter:1979bn, Antusch:2006vwa, Abada:2007ux}.

This model is MLFV because one can fully reconstruct the neutrino
Yukawa coupling $Y$ and the combination
$\widehat {Y'}=Y'-\frac{1}{\epsilon}\frac{\mu}{2\Lambda} Y$ from the
neutrino mass matrix~\cite{Gavela:2009cd} (up to two real
normalization factors $y$ and $\hat y'$) as follows:
\begin{equation}
  \begin{array}{c|c}
    {\rm NO}\; (m_1=0<m_2<m_3) & {\rm IO}\; (m_3=0 < m_1< m_2) \\[+0.3cm]
    \hline 
    r=\frac{\Delta m^2_{21}}{\Delta m^2_{32}} &
    r=-\frac{\Delta m^2_{21}}{\Delta m^2_{31}} \\[+0.3cm]
    \rho=\frac{\sqrt{1+r}-\sqrt{r}}{\sqrt{1+r}+\sqrt{r}} & 
    \rho=\frac{\sqrt{1+r}-1}{\sqrt{1+r}+1} \\[+0.3cm]
    m_{2,3}=\frac{\epsilon y {\widehat {y'}} v^2}{\Lambda}(1\mp\rho)
    &m_{1,2}=\frac{\epsilon y {\widehat {y'}} v^2}{\Lambda}(1\mp\rho)\\[+0.3cm]
Y_{a}=\frac{y}{\sqrt{2}}\left(\sqrt{1+\rho}\ U_{a3}^*+\sqrt{1-\rho}\ 
U_{a2}^*\right)  &
Y_{a}=\frac{y}{\sqrt{2}}\left(\sqrt{1+\rho}\ U_{a2}^*+\sqrt{1-\rho}\ 
U_{a1}^*\right) \\[+0.3cm]
\widehat {Y^\prime}_{a}= \frac
{\widehat{y^\prime}}{\sqrt{2}}\left(\sqrt{1+\rho}\ U_{a3}^*-
\sqrt{1-\rho}\ U_{a2}^*\right)  &
\widehat {Y^\prime}_{a}=
\frac{\widehat {y^\prime}}{\sqrt{2}}\left(\sqrt{1+\rho}\ U_{a2}^*-
\sqrt{1-\rho}\ U_{a1}^*\right)\\
  \end{array}
  \label{eq:reconsyuk}
\end{equation}

We plot in Fig.~\ref{fig:yuk} the ranges of the Yukawa couplings
$|\tilde Y_e|^2\equiv |Y_1|^2/y^2$ and
$|\tilde Y_\mu|^2\equiv |Y_2|^2/y^2$ obtained by projecting the
allowed ranges of oscillation parameters from the global analysis of
neutrino data \cite{Esteban:2016qun} using Eqs.~\eqref{eq:reconsyuk}.
In the first and second rows we plot the ranges of the Yukawa
couplings as a function of the unknown Majorana phase $\alpha$ while
the lower row shows the correlation between the electron and muon
Yukawa couplings.  This figure illustrates the quite different allowed
ranges and correlation of the electron and muon Yukawa couplings in
the two orderings.  As a curiosity we notice that in NO the dependence
of the range of $|Y_e|$ on $\alpha$ is driven by the present hint of
$\delta_{\rm CP}\sim 270^\circ$ in the oscillation data analysis
\cite{Esteban:2016qun} because $\alpha$ enters via
${\rm Re}(U_{e3} U_{e2}^*)\propto \cos(\alpha+\delta_{\rm CP})$.

In what respects the interactions of the heavy states, as discussed in
Ref.~\cite{Eboli:2011ia}, lepton number violating couplings appear due
to ${\cal O} (\epsilon, \mu/\Lambda, \mu^\prime/\Lambda)$ mixings and
mass splittings in the heavy states, that are, by hypothesis, very
suppressed in MLFV models.  This renders the lepton number violating
processes involving heavy fermions unobservable at LHC, at a
difference with the non MLFV scenarios for Type--III see--saw for
which $\Delta L=2$ final states constitute a smoking
gun~\cite{Franceschini:2008pz,Bajc:2007zf}.  Consequently in what
follows we concentrate in the lepton conserving interaction Lagrangian
with $\Lambda=M$ being the common mass of the heavy states:
\begin{eqnarray}
{\mathcal L}_W=&&-g\left(\overline{E^-_1}\gamma^\mu N W_\mu^- -
\overline{N}\gamma^\mu E^+_2W_\mu^-\right)+h.c. \nonumber \\
&&- g \left(\frac{1}{\sqrt{2}} K_{a}
\overline{\ell_{aL}}\gamma^\mu N_LW_\mu^- +
\tilde K_{a}\overline{\nu_{aL}} \gamma^\mu E^+_{2L} W_\mu^- \right) +h.c.
\label{eq:lw}
\\
{\mathcal L}_Z=&&gC_W
\left(
\overline{E^-_1}\gamma^\mu E^-_1Z_\mu-
\overline{E^+_2}\gamma^\mu E^+_2Z_\mu\right) \nonumber \\
&&+\frac{g}{2 \sqrt{2} C_W}\left(\frac{1}
         {\sqrt{2}}\tilde K_{a}\overline{\nu_{aL}}\gamma^\mu N_{L} Z_\mu
+ K_{a}\overline{\ell_{aL}}\gamma^\mu E^-_{1L}Z_\mu \right)+h.c.
\label{eq:lz}\\
{\mathcal L}_\gamma=&& e
\left(
\overline{E^-_1}\gamma^\mu E^-_1A_\mu-
\overline{E^+_2}\gamma^\mu E^+_2A_\mu\right) \\
{\mathcal L}_{h^0}=&&\frac{g M}{\sqrt{2}M_W}\left(
         \frac{1}{\sqrt{2}}\tilde K_{a}
\overline{\nu_{aL}} N_{R} +K_{a}\overline{\ell_{aL}}E^-_{1R}\right)+h.c. \; .
\label{eq:lh}
\end{eqnarray}
where $c_W$ stands for the cosine of the weak mixing angle and the
lepton number conserving couplings of the heavy triplet fermions are
\begin{equation}
\begin{array}{ll}
  K_a
  =-\frac{v}{\sqrt{2} M}  {Y_a}^{*} \;\;\; ,&\;\;\;  
  \tilde K_{a}={U^*_{LEP}}_{ca}  K_{c} \; ,
\end{array}
\label{eq:kdef}
\end{equation}
which verify
\begin{eqnarray}
\sum_{a=1}^3|K_a|^2=\sum_{a=1}^3|\tilde K_a|^2=\frac{y^2 v^2}{2 M^2} 
\; .
\label{eq:sumk}
\end{eqnarray}
Notice that flavour structures of the couplings $K$ and $\tilde K$ are
fully determined by the low energy neutrino parameters.  Their
strengths are, however, arbitrary as they are controlled by the
normalization factor $y v/M$ while it is the combination
$\epsilon y \widehat {y'}/M$ what is fixed by the neutrino masses.

From these interactions we can obtain the decay widths of the heavy
states~\cite{delAguila:2008hw}:
\begin{eqnarray}
&&\Gamma\left(N\rightarrow\ell_a^-W^+\right)=\frac{g^2}{64\pi}
|K_a|^2\frac{M^3}{M_W^2}
\left(1-\frac{M_W^2}{M^2}\right)\left(1+\frac{M_W^2}
{M^2}-2\frac{M_W^4}{M^4}\right) \; ,
\nonumber\\
&&\Gamma\left(N\rightarrow\nu_a Z\right)=\frac{g^2}{128\pi
c_w^2}|\tilde K_a|^2\frac{M^3}{M_Z^2}
\left(1-\frac{M_Z^2}{M^2}\right)\left(1+\frac{M_Z^2}{M^2}
-2\frac{M_Z^4}{M^4}\right) \; ,
\nonumber\\
&&\Gamma\left(N\rightarrow\nu_a
h^0\right)=\frac{g^2}{128\pi}|\tilde K_a|^2\frac{M^3}{M_W^2}
\left(1-\frac{M_{h^0}^2}{M^2}\right)^2 \; ,
\label{widths}\\
&&\Gamma\left(E_2^+\rightarrow\nu_a
W^+\right)=\frac{g^2}{32\pi}|\tilde K_a|^2\frac{M^3}{M_W^2}
\left(1-\frac{M_W^2}{M^2}\right)\left(1+\frac{M_W^2}{M^2}
-2\frac{M_W^4}{M^4}\right) \; ,
\nonumber\\
&&\Gamma\left(E_1^-\rightarrow\ell_a^-Z\right)=\frac{g^2}{64\pi
c_W^2}|K_a|^2\frac{M^3}{M_Z^2}
\left(1-\frac{M_Z^2}{M^2}\right)\left(1+\frac{M_Z^2}{M^2}
-2\frac{M_Z^4}{M^4}\right) \; ,
\nonumber\\
&&\Gamma\left(E_1^-\rightarrow\ell_a^-h^0\right)=\frac{g^2}{64\pi}
| K_a|^2\frac{M^3}{M_W^2}
\left(1-\frac{M_{h^0}^2}{M^2}\right)^2 \; .
\nonumber
\end{eqnarray}
Therefore, using Eq.~\eqref{eq:sumk} the total decay widths for the
three triplet fermions $F=N,E^-_1,E^+_2$ are
 \begin{eqnarray}
&&\Gamma_{F}^{\mbox{\begin{tiny}TOT\end{tiny}}}
=\frac{g^2 M^3}{64 \pi M_W^2} \frac{y^2 v^2}{M^2} 
(1+{\cal F}_F(M))  
\label{totalwidths}
\end{eqnarray}
where ${\cal F}_{F}(M)\rightarrow 0$ for $M\gg m_{h^0},M_Z,M_W$.
Consequently, the arbitrary $y$ factor cancels out in the branching
ratios.  Furthermore the branching ratio in a final state with a
charge lepton (neutrino) of flavour $\alpha$ produced in the vertex of
the heavy state decay is proportional to $|\tilde Y_\alpha|^2$
($|(U_{LEP} Y)_\alpha|^2$) times a kinematic factor depending solely
on $M$.

Unlike in a general Type--III see--saw model for which the branching
ratio of $N$ or $E^\pm_i$ into a light lepton of a given flavour can
be negligibly small, for the MLFV Type--III see--saw model the
branching ratios in the different light lepton flavours are fixed by
the neutrino physics and are non-vanishing as can be seen from
Fig.~\ref{fig:yuk} and Eqs.~(\ref{widths}).  This makes the flavour
composition of the final states in any decay chain of the heavy
leptons to be fully determined given a neutrino mass ordering.
Consequently in the narrow width approximation the only free
parameters in the model are the mass of the states and the Majorana
phase $\alpha$, making the model more unambiguously testable.
Conversely, as discussed in Ref.~\cite{Eboli:2011ia} in this MLFV
Type-III model the values of the neutrino masses imply a lower bound
on the total decay width of the triplet fermions as a consequence of
the hierarchy between the $L$-conserving and $L$-violating $y$ and
$\epsilon \widehat{y'}$ constants. So their decay length is too short
to produce a detectable displaced decay vertex
signature~\cite{Aad:2008zzm,Chatrchyan:2008zzk} at difference with
other see--saw models~\cite{Franceschini:2008pz, Bajc:2007zf,
  Perez:2008ha, Li:2009mw}.

In order to simulate the expected signals in this MLFV Type--III
see--saw model we have implemented the Lagrangian in
Eqs.~\eqref{eq:lw}--\eqref{eq:lh} using the package
FeynRules~\cite{Christensen:2008py,Alloul:2013bka}.
We have made available the corresponding model files  
at the corresponding URL~\cite{frwiki}.

\section{Case study: $pp\rightarrow l l' j j \nu\nu$}
\label{sec:simulation}

In order to study the sensitivity of LHC Run I to MLFV Type--III
see--saw signatures we will use the event topologies studied by ATLAS
in Ref.~\cite{Aad:2015cxa} which contain two charged leptons (either
electron or muons), two jets from a hadronically decaying $W$ boson
and large missing transverse momentum. ATLAS used these topologies to
search for heavy fermions in the context of the simplified Type--III
see--saw model as implemented in Ref.~\cite{Biggio:2011ja}.  They
presented their results as number of events for the six different
flavour and charge lepton pair combinations: same sign (SS) and
opposite sign (SS) $ee$, $\mu\mu$ and $e\mu$. Using those they obtain
bounds on the triplet mass which depend on the decay branching ratio
into the different flavours.  In particular for triplets decaying
mostly into $\tau$'s no bound can be derived.

On the contrary, as stressed in the previous section, in the MLFV
scenario the flavour and lepton number of the final states produced in
the heavy fermion decay chain is very much constrained.  Thus having
the final states classified in the different flavour and charge
combinations makes the result in Ref.~\cite{Aad:2015cxa} best suited
for testing the MLFV scenario as we quantify next.

\subsection{Contributing subprocesses}

Let us start by listing the possible subprocesses contributing to the
different flavour and charge combinations in the MLFV Type--III
see--saw model. They all proceed by the production of a pair of
heavy triplet states and their subsequent decay.
The pair production of the fermion triplets takes place via gauge
interactions, and, therefore, it depends exclusively upon the mass of the new
states. On the other hand, the branching ratios of these fermions into
the final states described in the previous section depend upon the
Yukawa couplings $\tilde{Y}_a$ which vary in the different subprocesses:
\begin{itemize}
\item 
$p \; p \; \rightarrow \; E_{1} ^- \; \tilde{N} \, , \quad
  ( \; E_{1}^- \; \rightarrow \; l_a ^- \;  Z \, , \, Z
  \;\rightarrow \;  \nu \; \nu  \; ) \quad
  ( \; \tilde{N} \; \rightarrow \; l_b ^+ \;  W^- \, , \, W^-   \;
  \rightarrow \;  j \; j  \;)$: \\[+0.3cm]
  Using interactions in \eqref{eq:lw}--\eqref{eq:lz} it is easy to
  show that the production cross section for this process is
  proportional to $\vert K_a \vert ^2 \, \vert K_b \vert ^2
  $. Therefore, as discussed in the previous section, in the narrow
  width approximation the production cross section for this process
  and its charge conjugated one can be factorized as
\begin{equation}
  \sigma_1(M)_{ab}\;
  \vert \tilde Y_a \vert ^2 \, \vert \tilde Y_b \vert ^2 .
  \label{eq:s1}
\end{equation}

\item $p \; p \; \rightarrow \; E_{2} ^+ \; \tilde{N} \,
 , \; ( \; E_{2}^+ \; \rightarrow \; \nu_c \;  W^+ \, , \,
  W^+ \;\rightarrow \;  j \; j  \; ) \quad
 ( \; \tilde{N} \; \rightarrow \; l_a ^+ \;  W^- \, ,
  \, W^-\; \rightarrow \;l_b^- \; \nu_c  \;)$: \\[+0.3cm]
  whose production cross section is proportional to
  $\vert K_a \vert^2 \vert \tilde{K}_c \vert^2$.  So in the narrow
  width approximation and after summing over the $\nu_c$ flavour using
  Eq.~\eqref{eq:sumk} the cross section for this process and its
  charge conjugated one can be written as
\begin{equation}
\sigma_{2a}(M)_{ab}\;\vert \tilde Y_a \vert^2\; .
\label{eq:s2a}
\end{equation}

\item
$p \; p \; \rightarrow \; E_{2} ^+ \; \tilde{N} \, ,
  \;  ( \; E_{2}^+ \; \rightarrow \; \nu_c \;  W^+ \, , 
  \, W^+   \; \rightarrow \;  l_b^+  \; \nu_b  \; ) \quad
  ( \; \tilde{N} \; \rightarrow \; l_a ^+ \;  W^- \, , 
  \, W^-   \; \rightarrow \; j \;\; j  \; )$: \\[+0.3cm]
  and the respective charge conjugate process possess a cross section
  proportional to $\vert K_a \vert^2 \vert \tilde{K}_c \vert^2$.  As
  before, in the narrow width approximation and after summing over the
  $\nu_c$ flavour we parametrize the production cross section as
\begin{equation}
  \sigma_{2b}(M)_{ab}\;\vert \tilde Y_a \vert^2\; .
\label{eq:s2b}
\end{equation}

\item
  $p \; p \; \rightarrow \; E_{2} ^+ \; \tilde{N} \, , \quad ( \;
  E_{2}^+ \; \rightarrow \; \nu_c \; W^+ \,\, , \,\, W^+ \;
  \rightarrow \; j \; j \; ) \quad ( \; \tilde{N} \;\rightarrow \;
  \nu_d \; Z \, ,
  \, Z   \; \rightarrow \; l_a^- \; l_a^+  \; ) $: \\[+0.3cm]
  and its charge conjugate process with cross section proportional to
  $\vert \tilde{K}_c \vert^2 \vert \tilde{K}_d \vert^2$ that after
  summing over the neutrino flavours $c$ and $d$ reads
\begin{equation}
\sigma_{2c}(M)_{aa} \;. 
\label{eq:s2c}
\end{equation}

\item
  $p \; p \; \rightarrow \; E_{1}^+ \; E_{1}^- \, , \quad
  ( \; E_{1}^+ \; \rightarrow \; l_a^+ \;  Z  \, , \, Z
  \; \rightarrow \;  j  \; j  \;) \quad 
  (\; E_1^- \; \rightarrow \; l_b^- \;   Z   \, , \,  Z
  \; \rightarrow \; \nu_c \; \nu_c  \;) $\, , \\
  $p \; p \; \rightarrow \; E_{1}^+ \; E_{1}^- \, , \quad
  ( \; E_{1}^+ \; \rightarrow \; l_a^+ \;  Z  \, , \,
  Z  \; \rightarrow \;  \nu_c \; \nu_c   \; ) \quad
  (\; E_1^- \; \rightarrow \; l_b^- \;  Z   \, , \,  Z
  \; \rightarrow \;  j \;j  \; ) $ : \\[+0.3cm]
with  cross section 
\begin{equation}
  \sigma_{3}(M)_{ab}\; \vert \tilde Y_a \vert^2
  \vert \tilde Y_b \vert^2 \; .
  \label{eq:s3}
\end{equation}
\end{itemize}

Altogether the cross section of each OS flavour channel is given by:
\begin{eqnarray}
  \sigma^{OS}_{ee}&\equiv&
 \left[ \sigma_1(M)_{ee}  \,
    +\, \sigma_3(M)_{ee}\right] \vert\tilde Y_e \vert^4 \,
  +\, \sigma_{2a}(M)_{ee} \vert \tilde Y_e \vert^2
  +\, \sigma_{2c}(M)_{ee} \; , \nonumber \\
  \sigma^{OS}_{e\mu}&\equiv &
  \left[ \sigma_1 (M)_{e\mu}   \,
    +\,\sigma_1(M)_{\mu e} \,
    +\, \sigma_3(M)_{e \mu}  \,
    +\,  \sigma_3(M)_{\mu e}   \right]
  \vert \tilde Y_e \vert^2 \vert \tilde Y_{\mu} \vert^2 \,\nonumber \\
 &&  +\, \sigma_{2a}(M)_{e \mu} \vert \tilde Y_e \vert^2
\,+\,\sigma_{2a}(M)_{\mu  e}
  \vert\tilde Y_{\mu} \vert^2 \; , \label{eq:OS}
\\
  \sigma^{OS}_{\mu \, \mu}&\equiv&
  \left[\sigma_1(M)_{\mu\mu}  \,+\,
    \sigma_3(M)_{\mu\mu}\right] \vert \tilde Y_{\mu} \vert^4 \,
  +\, \sigma_{2a}(M)_{\mu\mu} \vert\tilde Y_{\mu} \vert^2
 \,+\, \sigma_{2c}(M)_{\mu\mu} \;,\nonumber
\end{eqnarray}
while for SS lepton final state
\begin{eqnarray}
 \sigma^{SS}_{ee}&\equiv& 
  \sigma_{2b}(M)_{ee} \vert \tilde Y_e \vert^2 \; , \nonumber \\
  \sigma^{SS}_{e \mu}&\equiv& \sigma_{2b}(M)_{e \mu}
  \vert \tilde Y_e \vert^2 
  +\, \sigma_{2b}(M)_{\mu e}  \vert\tilde Y_{\mu} \vert^2 \; , \label{eq:SS}
 \\
  \sigma^{SS}_{\mu\mu}&\equiv & \sigma_{2b}(M)_{\mu\mu}
  \vert\tilde Y_{\mu}\vert^2 \; . \nonumber 
\end{eqnarray}

\subsection{Simulation of the expected event rates}

In our analysis, we first simulated the above parton level signal processes
with MadGraph 5~\cite{Alwall:2014hca}.  We then used
PYTHIA 6.4 ~\cite{Sjostrand:2006za} to generate the parton shower and
hadronization.  Finally we performed a fast detector simulation with DELPHES
3~\cite{deFavereau:2013fsa} with jets being reconstructed using the
anti-$k_T$ algorithm with a radius $R=0.4$ with the package
FASTJET~\cite{Cacciari:2011ma}.

In order to reproduce the ATLAS event selection corresponding to the
searches in Ref.~\cite{Aad:2015cxa} we required that the events
contain exactly two leptons (muons and/or electrons), a minimum of two
jets and no b-tagged jet. In the case of OS (SS) leptons the leading
lepton must have transverse momentum ($p_T$) in excess of 100 (70) GeV
with the next-to-leading lepton $p_T$ larger than 25 (40) GeV. In
addition we imposed the invariant mass of the two leptons to be larger
than 130 and 90 GeV for OS and SS events respectively.  We also
demanded the $p_T$ of two leading jets to be larger than 60 (40) and
30 (25) GeV for the OS (SS) final state. Moreover, to characterize the
presence of a hadronically decaying $W$ in the event the invariant
mass of the leading jet pair was required to be between 60 and 100
GeV, and for OS events we require the two leading jets to satisfy
$\Delta R_{jj} < 2$.  Finally, we only selected events presenting
missing transverse energy in excess of 110 (100) GeV for OS (SS)
events.  We implement the above selection in
MadAnalysis5~\cite{Conte:2012fm}.

In order to tune our calculations we first simulated the signal for the
Type-III see--saw model of Ref.~\cite{Biggio:2011ja} which is the one
used in the ATLAS analysis. By comparing our number of expected events with
the ones obtained by ATLAS in Fig. 2 of~\cite{Aad:2015cxa} for the
OS and SS final states presenting $ee$, $e\mu$ and $\mu\mu$ pairs, we 
extract overall multiplicative correction factors for each of  these final states
such that our fast simulation agrees with the one of ATLAS.
To validate our procedure we 
verified that our tuned Monte Carlo reproduces the missing transverse
momentum distribution presented in Fig. 1 of~\cite{Aad:2015cxa}.
We then apply these  correction factors in the evaluation of the expected number
of events of the MLFV type-III see-saw model.

Concerning backgrounds, the dominant contribution comes from the
production of dibosons ($WW$,$WZ$,$ZZ$), $Z$ plus jets, top pairs and
single top in association with a $W$. In addition there are also
background events that stems from misidentification of leptons. In our
analysis we directly use the background rates estimated by
ATLAS~\cite{Aad:2015cxa}.

\begin{figure}\centering
\includegraphics[width=0.8\textwidth]{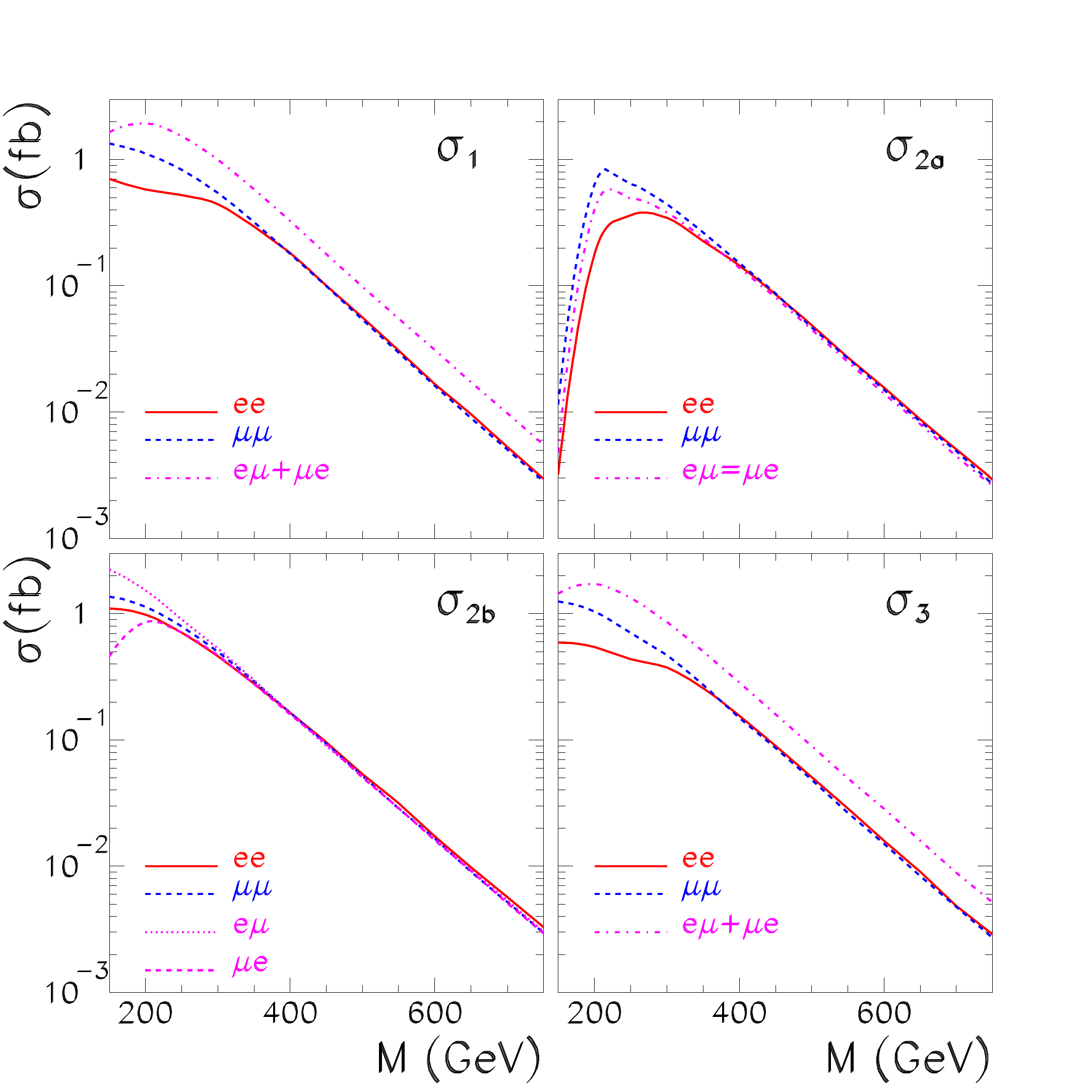}
\caption{Cross section factors for the different contributions
  to the event topologies as a function of the triplet mass $M$
  as defined in Eqs.~\eqref{eq:s1}--\eqref{eq:s3} after inclusion of our
  simulation of the selections of Ref.~\cite{Aad:2015cxa} (see text for
  details).}
  \label{fig:sigmas}
\end{figure}

Figure~\ref{fig:sigmas} depicts the resulting cross section factors
$\sigma_1$, $\sigma_{2a}$, $\sigma_{2b}$, and $\sigma_3$ introduced in 
Eqs.~(\ref{eq:s1})--(\ref{eq:s3}),  for the  events passing the selection cuts and
after applying the tuning correction factors. The results are shown  as a
function of the new fermion mass and for the different lepton flavor
combinations and a center-of-mass energies of 8 TeV~\footnote{$\sigma_{2c}$
is vanishing small since the event selection vetoes the presence of on-shell
$Z$'s.} . From this figure
we can see that the four possible leptonic final states ($ee$,
$\mu\mu$, $e\mu$, and $\mu e$) have similar cross section factors for
masses larger than 300 GeV, however, closer to the threshold the
detection and acceptance efficiencies induce differences among the leptonic
final states. In the case of $\sigma_{2a}$ both leptons originate from
the decay of a single triplet fermion, therefore reducing the
acceptance at small masses due to the OS lepton pair invariant mass
cut.

\begin{figure}\centering
\includegraphics[width=0.8\textwidth]{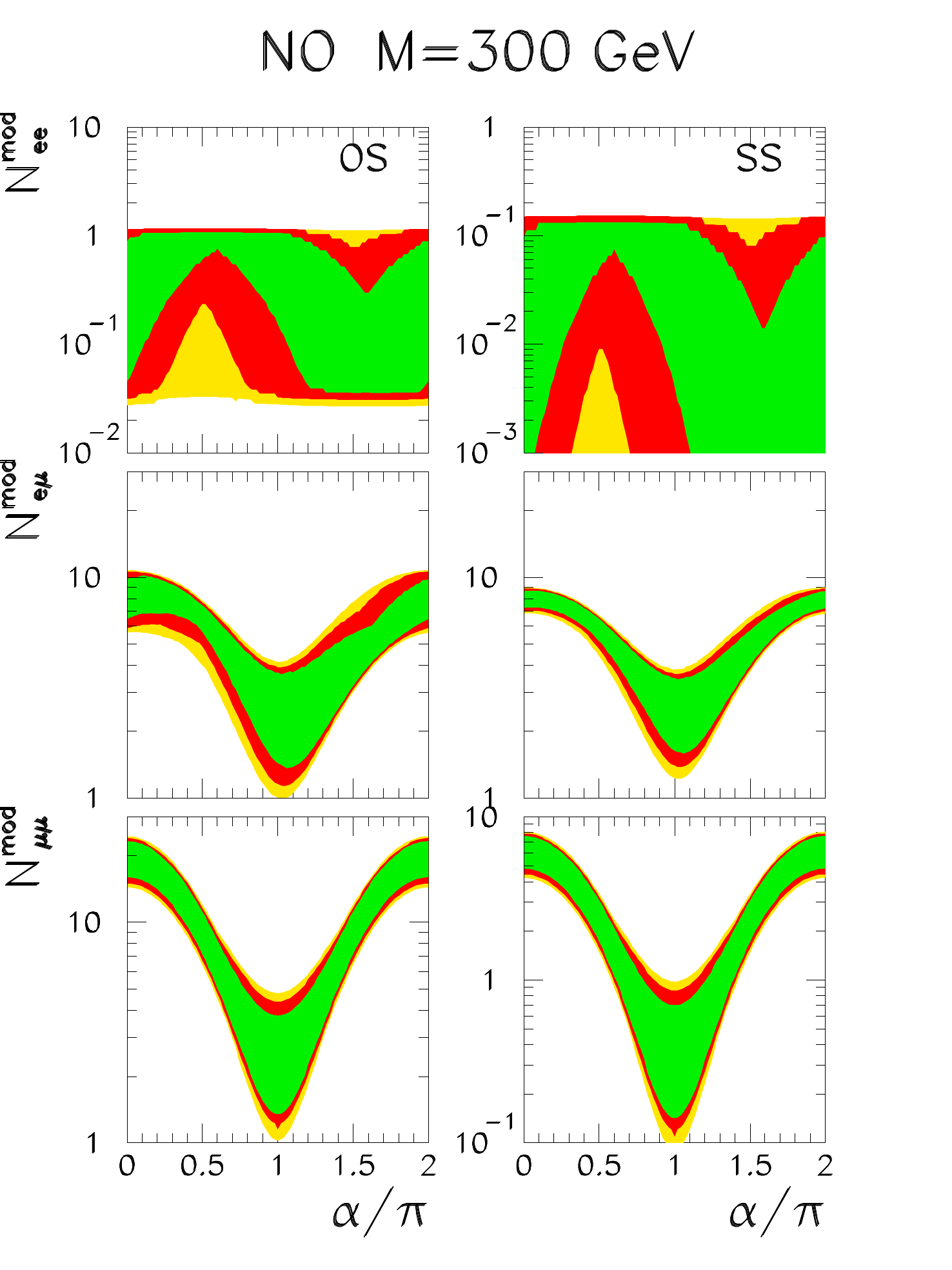}
\caption{Expected number of signal events at $\sqrt{s}=8$ TeV
  and an integrated luminosity of 20.3 fb$^{-1}$,
  in each of the six flavour-charge
  combinations for a triplet mass of $M=300$ GeV  
  and for neutrino parameters corresponding to the NO.
  The conventions are  the same as in Fig.1.}
\label{fig:neventsno}
\end{figure}

\begin{figure} [h]
  \centering
\includegraphics[width=0.8\textwidth]{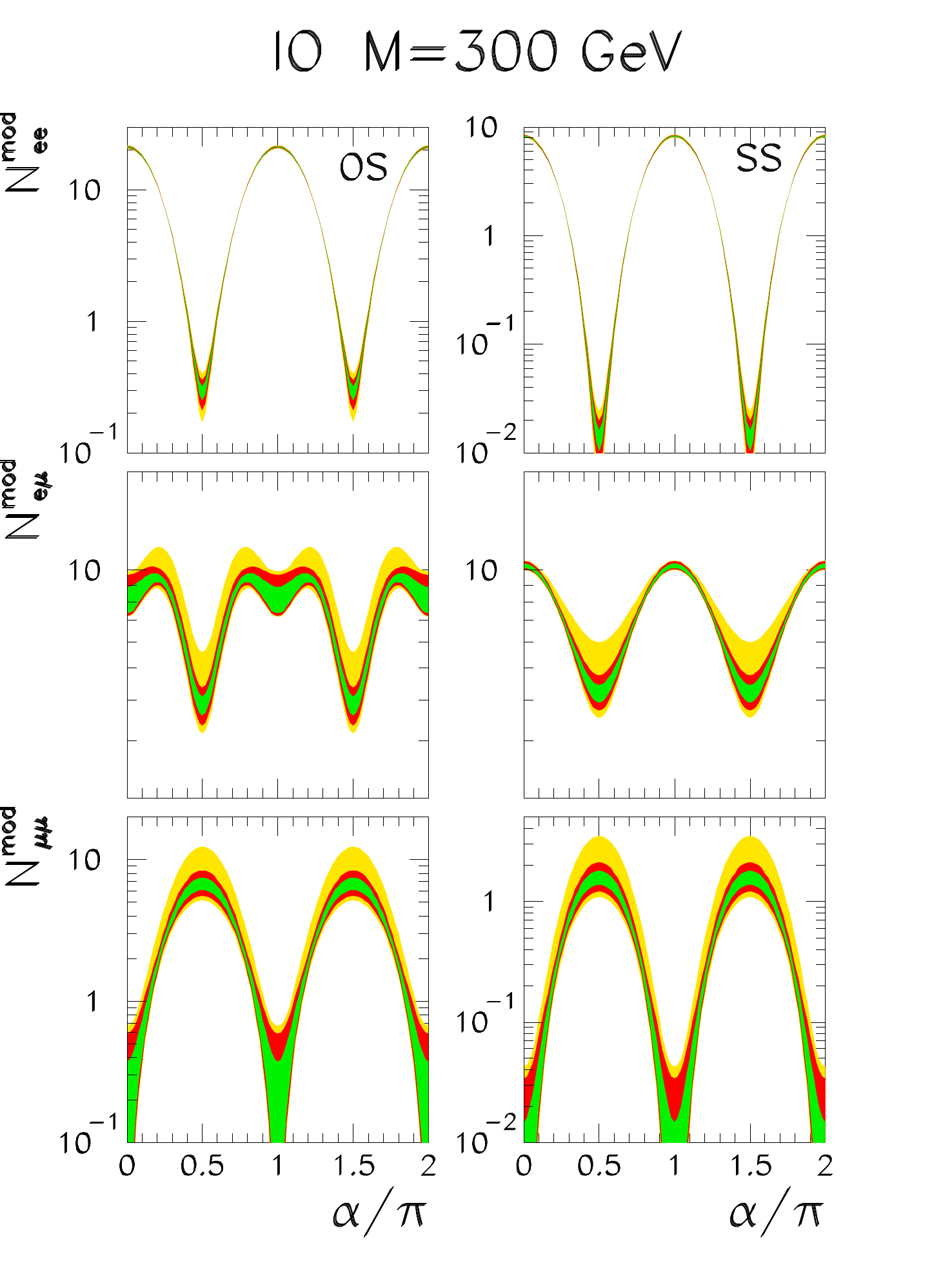}
\caption{Same as Fig.~\ref{fig:neventsio} for  IO.}
\label{fig:neventsio}
\end{figure}

Weighting these cross section factors with the different combination of
Yukawa couplings as in Eqs.~\eqref{eq:OS} and ~\eqref{eq:SS} and times the
luminosity we predict the signal event rates in the different flavour and charge
combinations of the final lepton pair.
For example in Figs.~\ref{fig:neventsno}
and ~\ref{fig:neventsio}  we present the event rate prediction as a function
of the unknown Majorana phase $\alpha$ for a triplet mass of 300 GeV,
an integrated luminosity of 20.3
fb$^{-1}$ and normal ordering and inverted ordering respectively.
As seen in Fig.~\ref{fig:neventsno} for NO, the expected number
of $ee$ events is very small for both OS and SS
channels. Nevertheless, a considerable number of events containing
muons is expected except for $\alpha/\pi$ around 1.  This is so because for NO
$|\tilde{Y}_\mu|$ is much larger than $| \tilde{Y}_e|$ as can be
seen from Fig.~\ref{fig:yuk}. For IO we see in Fig.~\ref{fig:neventsio},
as could be anticipated from Fig.~\ref{fig:yuk}, that 
the expectations in all flavor channels vary appreciably with the Majorana
phase $\alpha$. However, the strong correlation between $|\tilde{Y}_e|$ and
$| \tilde{Y}_\mu|$ in IO -- depicted in the right bottom panel of
Fig.~\ref{fig:yuk} -- guarantees a sizable number of events for almost
the entire $\alpha$ range with the signal being dominated by the
channels $e\mu$ and $\mu\mu$ for OS leptons and the channel $e\mu$ for
SS leptons. 

\section{Analysis: Results and Discussion}
\label{sec:results}

In order to quantify the bounds on the MLFV Type--III see--saw
scenario we build the likelihood function using the six data points
associated to the events with $ee$, $e\mu$ and $\mu\mu$ leptons of
either SS or OS. As discussed in the previous section, we make direct use 
of the corresponding background
estimate by ATLAS which we read from Fig. 2 and Table I of
Ref.~\cite{Aad:2015cxa} and that summarize here for completeness:
\vskip 0.5cm
\begin{tabular}{c|cccccc|cc}
  & OS $ee$ & SS $ee$ & OS $\mu\mu$ & SS $\mu\mu$ & OS $e\mu$ & SS $e\mu$
  & total OS & total SS \\
  \hline
  $N^{dat}$ & 9 & 3 & 3 &1 & 13 & 0 & 25 &4  \\
  $ N^{bck}$ & $8.5$ & $1$ & $9.5$ & $0.5$
  & $13$
  & $1.65$ & $31.0\pm 7.7$& $3.15\pm 0.8$ 
\end{tabular}
\vskip 0.5cm
According to Ref.~\cite{Aad:2015cxa} the reported background errors in
the table and figure include both the statistical and systematic
uncertainties.  Comparing the read out of these uncertainties for each
of the six individual channels with the total reported in the table we
conclude that the $\sim$ 25\% background uncertainty is strongly
correlated among the different channels as the total background
uncertainty comes to be very close to the arithmetic sum of the
individual ones, while if they were totally uncorrelated one would
expect it to be the quadratic sum.

So we build the likelihood function as
\begin{equation}
  -2{\cal L}_{6d}=\chi^2_{6d}  =\min_{\xi}
  \left\{ 2\sum_{i=1,6}\left[ (1+\xi)N^{bck}_i
    +N^{mod}_i-N^{dat}_i \log\frac{N^{dat}_i}{(N^{bck}_i+N^{mod}_i)}\right]
  +\frac{\xi^2}{0.25^2} \right \}
  \label{eq:l6}
\end{equation}  
where we account for the background uncertainty by introducing a
unique pull $\xi$  with an uncertainty of 25\%\footnote{ We have
  verified that including several pulls for the different source of
  background and smaller uncertainties each does not have any
  significant impact in the results.}. $N^{mod}_i$ is the predicted
contribution to the number of events in channel $i$ from the triplet
production and decays which depends on the triplet mass and the
neutrino parameters as discussed in the previous section. In building
the likelihood \eqref{eq:l6} we have used Poisson statistics to
account for the small number of events in each channel.

\begin{figure} [h]
  \centering
\includegraphics[width=0.6\textwidth]{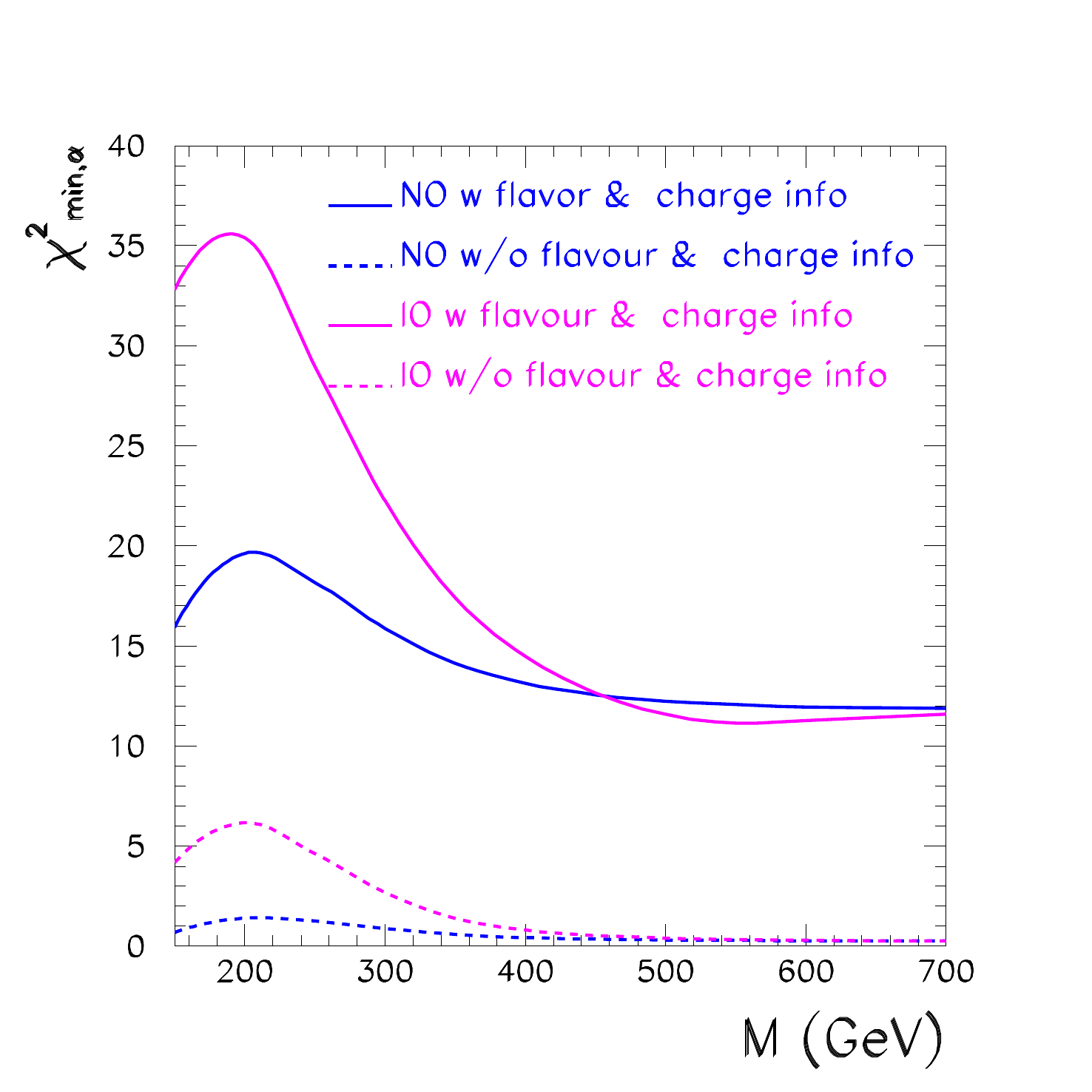}
\caption{Triplet mass dependence of the $\chi^2$ functions of the analysis of
  $pp\rightarrow l l' j j \nu\nu$  events with $l,l'$ being either
  $e$ or $\mu$ of either charge observed in the LHC Run-I in
  ATLAS~\cite{Aad:2015cxa} when analyzed in the context of MLFV Type--III
  see--saw model. The full lines correspond to the likelihood
  function constructed including the full information
  given on flavour and charge of the final states (see Eq.~\eqref{eq:l6})
  while the dashed lines are obtained from the analysis of the total
  data summing over charge and flavour. The triplet couplings have been
  marginalized within the ranges allowed at 95\% CL by the neutrino oscillation
  data analysis in Ref.~\cite{Esteban:2016qun} for NO (blue lines)
  and IO (purple lines) and for any value of the Majorana phase $\alpha$.}
  \label{fig:chi2min}
\end{figure}

We plot as full lines in Fig.~\ref{fig:chi2min} the dependence of
$\chi^2_{6d}$ on the triplet mass after marginalization over the
neutrino parameters (including the unknown Majorana phase $\alpha$)
over the 95\% CL allowed values from the neutrino oscillation analysis
for either NO or IO.  First thing to notice is that in the SM
($N^{mod}_i$=0, which can be inferred from the large M limit in the
figure) we find $\chi^2_{6d,SM}=11.8$ which is a bit high for 6 data
points. This is mostly driven by the OS $\mu\mu$ channel for which 3
events are observed when about 10 are expected in the SM.  From this
figure we also read that requiring $\chi^2_{6d}-\chi^2_{6d,SM}<4$ we
can infer an absolute bound on the triplet mass of 300 GeV (375 GeV)
for NO (IO) light neutrino masses.
 
In order to stress the importance of the flavour and charge
information on the possibility of imposing this bound, we have
constructed the corresponding likelihood function summing the
information from all the channels. As in this case the total number of
observed events is large enough, we assume gaussianity.  So we define:
\begin{equation}
\chi^2_{tot}=\frac{(N^{tot,bck}+N^{tot,mod}-N^{tot,dat})^2}
{N^{tot,dat}+(0.25 N^{tot,bck})^2}  \; .
\label{eq:l1}
\end{equation}  
The dependence of $\chi^2_{tot}$ on the triplet mass after
marginalization over the neutrino parameters (including the unknown
Majorana phase $\alpha$) over the 95\% CL allowed values from the
neutrino oscillation analysis ($\Delta\chi^2_{osc}\leq 4$) for either
NO or IO is shown as dashed lines in Fig.~\ref{fig:chi2min}.  As the
deficit in OS $\mu\mu$ is compensated by the slight excesses in other
channels, we find that in this case the SM gives a perfect description
of the total observed event rates ($\chi^2_{tot,SM}$=0.25). The figure
clearly illustrates the relevance of flavour and charge information,
as in this case the condition $\chi^2_{tot}-\chi^2_{tot,SM}<4$ does
result into no bound on the triplet mass for NO neutrino masses while
it rules out only $M<260$ for IO.

\begin{figure} [h]
  \centering
\includegraphics[width=0.8\textwidth]{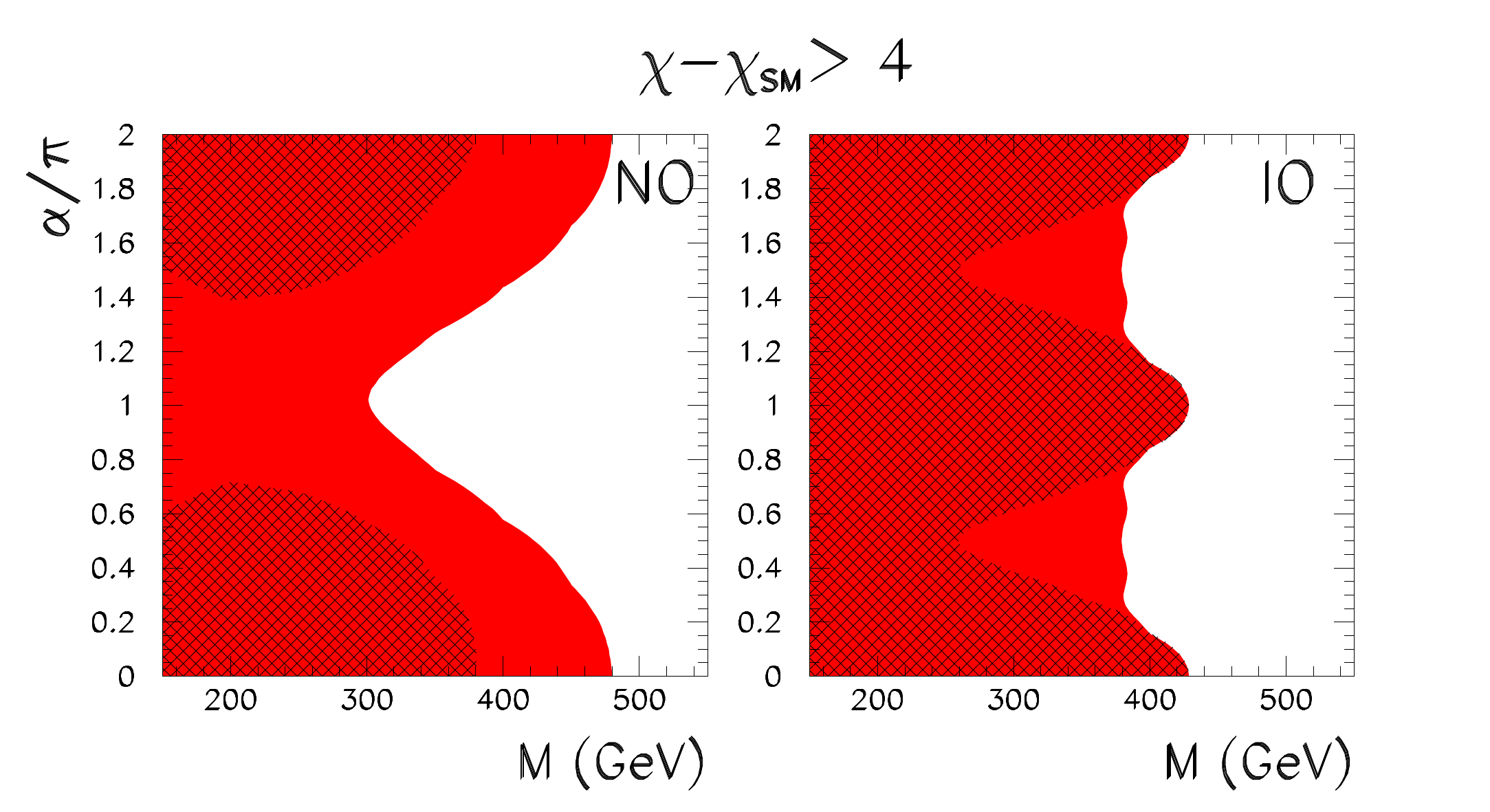}
\caption{95\% excluded triplet mass in the MLFV Type--III see--saw scenario 
  as a function of the unknown phase $\alpha$ from the analysis of
  $pp\rightarrow l l' j j \nu\nu$  events with $l,l'$ being either
  $e$ or $\mu$ of either charge observed in the LHC Run-I in
  ATLAS~\cite{Aad:2015cxa}. The full regions correspond to the likelihood
  function constructed including the full information
  given on flavour and charge of the final states (see Eq.~\eqref{eq:l6})
  while the hatched ones are obtained from the analysis of the total
  data summing over charge and flavour (see Eq.~\eqref{eq:l1}).
  The triplet couplings have been
  marginalized within the ranges allowed at 95\% CL by the neutrino oscillation
  data analysis in Ref.~\cite{Esteban:2016qun} for NO (left)
  and IO (right).}
  \label{fig:bounds}
\end{figure}

The dependence of the bounds on the unknown phase $\alpha$ is
displayed in Fig.~\ref{fig:bounds}. The full red regions are excluded
values of triplet masses in the MLFV scenario at 95\% CL
($\chi^2_{6d}-\chi^2_{6d,SM}>4$) when marginalizing over the
oscillation parameters within the 95\% CL allowed values by the
oscillation analysis in NO (left) and IO (right) for each value of
$\alpha$.  The $\alpha$ marginalized bound discussed above correspond
to the lightest allowed masses in these panels which for NO ($M>300$
GeV) correspond to $\alpha=\pi$ while for $\alpha=0$, $2\pi$ the bound
strengthens to $M>480$ GeV.  For IO the dependence of the bound on
$\alpha$ is weaker. The marginalized bound $M>380$ GeV
corresponds to $\alpha\sim 3\pi/4,3\pi/2$ but it is close to that value for
almost all values of alpha.  The strongest bound of
$M>430$ GeV corresponds also to $\alpha=0$, $2\pi$.

The hatched regions are the corresponding constraints obtained using
only the information on the total number of events
($\chi^2_{tot}-\chi^2_{tot,SM}>4$) summed over flavour and charge of
the final leptons.  This figure illustrates again how using the
flavour and charge information allows to impose stronger bounds on
this scenario, in particular allowing to rule out triplet masses
irrespective of $\alpha$ for both orderings while for NO no bound can
be imposed for $70^\circ \pi\lesssim \alpha \lesssim 250^\circ$ if
only the total number of events is considered.

In summary, we have shown how the analysis of the events containing
two charged leptons (either electron or muons), two jets from a
hadronically decaying $W$ boson in the Run I with the ATLAS detector
~\cite{Aad:2015cxa} can be used to impose constraints on the MLFV Type
III see--saw scenario. Because of MLFV, the expected event rates in
the different flavour and charge combinations of the two leptons are
constrained by the existing neutrino data so the bounds cannot be
evaded.  For this reason it is possible to use this data to rule out
these scenario with triplet masses lighter than 300 GeV at 95\% CL
irrespective of the neutrino mass ordering and of the value of the
unknown Majorana phase parameter. The same analysis allows to rule out
triplet masses masses up to 480 GeV at 95\% CL for NO and
$\alpha=0,\pi$.  We have stressed and quantified how the information
on flavour and charge information of the produced leptons is important
for maximal sensitivity to MLFV.

We finish by commenting that extended sensitivity to MLFV with heavier
triplets should be attainable with the data already accumulated from
Run II in the same or other event topologies. For example by the
analysis of the multilepton final states in CMS in
Ref.~\cite{CMS:2017wua} which, so far, has been performed only in the
context of the simplified Type--III see--saw model.  Nevertheless, as
previously stressed to do so it is important to make use of the
flavour and charge of final state leptons which has not been made
public yet. To this aim, we have made available the  model files
for the MLFV Type-III see-saw~\cite{frwiki}.

\acknowledgments M.C.G-G wants to thank her NUFIT collaborators,
I. Esteban, M. Maltoni, I. Martinez and T. Schwetz for their generous
contribution of the results from the data oscillation analysis used in
this article. She also wants to thank the USP group for their their
hospitality during the final stages of this work. This work is
supported in part by Conselho Nacional de Desenvolvimento
Cient\'{\i}fico e Tecnol\'ogico (CNPq) and by Funda\c{c}\~ao de Amparo
\`a Pesquisa do Estado de S\~ao Paulo (FAPESP) grants 2012/10095-7 and
2017/06109-5, by USA-NSF grant PHY-1620628, by EU Networks FP10 ITN
ELUSIVES (H2020-MSCA-ITN-2015-674896) and INVISIBLES-PLUS
(H2020-MSCA-RISE-2015-690575), by MINECO grant FPA2016-76005-C2-1-P
and by Maria de Maetzu program grant MDM-2014-0367 of ICCUB.

\bibliographystyle{JHEP}
\bibliography{references}

\end{document}